\documentstyle[12pt,amstex,amssymb] {article}
\newcommand{\beq}{\begin{equation}}

\newcommand{\eeq}{\end{equation}}
\newcommand{\beqa}{\begin{eqnarray}}
\newcommand{\eeqa}{\end{eqnarray}}
\begin {document}
\parindent=15pt
\begin{flushright}
{\bf US-FT/7-97}
\end{flushright}
\vskip .8 truecm
\begin{center}
{\bf MULTIPLICITY AND TRANSVERSE ENERGY DISTRIBUTIONS ASSOCIATED TO 
RARE EVENTS IN NUCLEUS-NUCLEUS COLLISIONS.}\\
\vskip 1.5 truecm
{\bf J. Dias de Deus$^*$, C. Pajares and C. A. Salgado.}\\
\vskip 0.9 truecm
{\it Departamento de Fisica de Part\'{\i}culas, Universidade de Santiago de
Compostela, \\
15706--Santiago de Compostela, Spain}
\end{center}
\vskip 2.5 truecm

\begin{abstract}
\hspace {30pt}

We show that in high energy nucleus-nucleus collisions the transverse energy or
multiplicity distribution $P_C$, associated to the production of a rare,
unabsorbed event C, is universally related to the standard or minimum bias
distribution $P$ by the equation 
$$
P_C(\nu)={\nu\over<\nu>}P(\nu)\ \ \ ,
$$

\noindent
with $\sum P(\nu)=1$ and $\nu\equiv E_T$ or $n$. Deviations from this formula
are discussed, in particular having in view the formation of the plasma of
quarks and gluons. This possibility can be distinguished from absortion or
interaction of comovers, looking at the curvature of the $J/\Psi$ over
Drell-Yan pairs as a function of $E_T$.
\end{abstract}

*) Iberdrola visiting professor. On leave of absence from 
Instituto Superior T\'ecnico, 1096 LISBOA codex, Portugal.

\vskip 1cm
February 1997 \\
\vskip .2 truecm

{\bf US-FT/7-97}

\pagebreak

It is clear, from theoretical models and experiment \cite{1,2}, that 
particle production in hadron-hadron, hadron-nucleus ans nucleus-nucleus
collision is generated by superposition of elementary, particle emitting, 
collisions. In the Dual Parton Model (DPM) \cite{3}, for instance, the 
elementary collisions involve the partonic constituents of hadrons and
the production of particle emitting strings.

Assuming that all the elementary collisions may be treated as equivalent,
particle production fluctuations will contain a contribution from the 
fluctuation in the number of elementary collisions (or the number of strings)
and from the fluctuation in particle production resulting from the elementary
collision (string particle distribution) \cite{4}. The fact that the width
of the nucleus-nucleus particle distribution is very large and very 
different from the width of the $e^+e^-$ distribution or even hadron-hadron
distribution, is an indication that multiparticle fluctuations in 
nucleus-nucleus  
collisions are dominated by fluctuations in the number
$\nu$ of elementary collisions.

In general, we then have for the average multiplicity $<n>$ and for the
square of the dispersion $D^2\equiv<n^2>-<n>^2$:

\begin{equation}
<n>=<\nu>\bar n \ \ , 
\label{1} 
\end{equation}

\noindent
and 
\begin{equation}
{D^2\over<n>^2}={<\nu^2>-<\nu>^2 \over <\nu>^2 }+{1 \over <\nu>}{d^2 \over 
\bar n^2} \ \ ,  
\label{2}
\end{equation}

\noindent
where $\nu$ and $d$ are the average multiplicity and the dispersion of the
elementary collision particle distribution, respectively.

In nucleus-nucleus high energy collisions, as the left hand side of (\ref{2})
is of the order of 1 and the second term in the right hand side is of the 
order of $10^{-2}-10^{-3}$, naturally a good approximation is obtained by
writing
\begin{equation}
{D^2\over<n>^2}={<\nu^2>-<\nu>^2\over <\nu>^2 } 
\label{2p}
\end{equation}

\noindent
The same kind of approximation is valid for higher moments \cite{4}.

In order to obtain (\ref{1}) and (\ref{2p}) we make the simplifying assumption

\begin{equation}
P(n)=P(\nu),\ \ \ \ \ n=\nu\bar n\ \ .
\label{3}
\end{equation}

Equation (\ref{3}) essentially tells us that, in nucleus-nucleus interactions,
the KNO distribution \cite{5} for particles is well approximated by the KNO 
distribution for elementary collisions.

Note that, as the transverse energy $E_T$ is a good measure of $n$, 
Eq.(\ref{3}),  
and what follows, can as well be written for $E_T$ distributions.

Let us next consider the particle distribution associated to a rare event C 
(Drell-Yan pairs in some mass region, $J/\Psi$ and $\Psi'$ production in
rough approximation, $\Upsilon$ production, some weak process, etc.) which
does not suffer strong absortion. These are events of type C in the 
classification of \cite{6}.

If $\alpha_C$, $0>\alpha_C>1$ is the probability of event C to occur in an 
elementary collision and if in a nucleus-nucleus experiment $N(\nu)$ is the
number of events with $\nu$ elementary collisions, we have

\begin{equation}
N(\nu)=\sum_{i=0}^{\nu} \binom{\nu}{i} (1-\alpha_C)^{\nu-i}
\alpha_C^i N(\nu)\ \ ,
\label{4}
\end{equation}

\noindent
where $(1-\alpha_C)^\nu$ is the probability of C not occuring, $\nu
(1-\alpha_C)^{\nu-1}\alpha_C$ the probability of C occuring once, etc. If event
C is rare we can approximate (\ref{4}) by the leading terms in $\alpha_C$ (this
is our definition of rare event):

$$
N(\nu)=(1-\alpha_C\nu)N(\nu)+\alpha_C\nu N(\nu)\ \ , 
$$
\noindent
where

\begin{equation}
N_C(\nu)=\alpha_C\nu N(\nu)
\label{5}
\end{equation}

\noindent
is the number of events where event C occurs. If $N$ is the total number of
events, we have

\begin{eqnarray}
\sum_{\nu}N(\nu)=N\ \ ,
\label{6} \\
\sum_{\nu}\nu N(\nu)=<\nu>N\ \ , \ \ etc.\ \ ,
\label{7}
\end{eqnarray}

\noindent
and, for the total number of events with C occuring
\begin{equation}
\sum_{\nu}(\alpha_C\nu)N(\nu)=\alpha_C<\nu>N\ \ .
\label{8}
\end{equation}
\noindent
This implies, for the probability distribution

\begin{equation}
P_C(\nu)\equiv{\alpha_C\nu N(\nu)\over\alpha_C<\nu> N}
={\nu\over <\nu>}P(\nu)\ \ .
\label{9}
\end{equation}

\noindent
Within approximation (\ref{3}) we finally obtain,

\begin{equation}
P_C(n)={n\over<n>}P(n)
\label{10}
\end{equation}
\noindent
and, equivalently,

\begin{equation}
P_C(E_T)={E_T\over <E_T>}P(E_T)\ \ .
\label{11}
\end{equation}
\noindent
Relations (\ref{10}) and (\ref{11}) are universal, independent of $\alpha_C$.

Our main result, (\ref{10}) and (\ref{11}), depends on assumptions to be
discussed now.

i) The first assumption is the dominance of fluctuactions in the number of
elementary collisions, as seen in (\ref{3}). This dominance increases with the
increase of $<\nu>$. This means going to heavier nuclei and higher energies.
Specific DPM calculations suggest that this approximation is quite reasonable
\cite{7}.

ii) The second assumption is the smallness of the probability of the events C
to occur. One can take (\ref{4}) to higher orders in $\alpha_C$, if $\alpha_C$
is not small enough. In that case one may still obtain a relation between
$P_C(n)$ and $P(n)$, but it will depend on $\alpha_C$: universality is lost.

iii) The third assumption is the assumption of linearity in (\ref{5}) of the
dependence of the probability of events C on $\nu$, for previous dicussion see
\cite{8}. Final state destructive absortion, as in $J/\Psi$ production, for
instance, will eliminate this linearity by making the effective number of
collisions where event C appears smaller. This can be visualized by making in
(\ref{5}) the change

\begin{equation}
\alpha_C\nu N(\nu)\longrightarrow \alpha_C\nu^\varepsilon N(\nu),\ \ \
\varepsilon<1  
\label{5p}
\end{equation}
\noindent
and

\begin{equation}
P_C(n)={n^\varepsilon\over<n^\varepsilon>} P(n)\ \ \ ,
\label{10p}
\end{equation}

\noindent
or
\begin{equation}
P_C(E_T)={E_T^\varepsilon\over<E_T^\varepsilon>} P(E_T)\ \ \ ,
\label{11p}
\end{equation}

\noindent
It is clear that, as $0<\varepsilon<1$, absortion makes the associated
distribution closer to the standard distribution.

We would like, at this stage, to check the validity of (\ref{11}) and
({\ref{11p}), by comparing Drell-Yan, $J/\Psi$ and minimum bias $E_T$
distributions of the S-U experiment of NA38 Collaboration \cite{9}.

Relation (\ref{11}), for Drell-Yan production in comparison with minimum bias
data is tested in Fig.1. The agreement is quite good. Best agreement, having in
mind absortion, (\ref{11p}), is obtained for $\varepsilon_{DY}\simeq 0.95$ (not
shown in the Figure). In Drell-Yan events absortion is not important.

In the case of $J/\Psi$ production, see Fig.2, the best agreement, from
(\ref{11p}), is obtained for $\varepsilon_{J/\Psi}\simeq 0.7$. In this case
absortion cannot be neglected.

One can also directly compare the $J/\Psi$ production to DY production. From
(\ref{5p}), with $\nu\equiv E_T$,

\begin{equation}
N_{J/\Psi}(E_T)/N_{DY}(E_T) \sim 1/E_T^\gamma\ \ \ ,
\label{12}
\end{equation}

\noindent
and $\gamma\equiv\varepsilon_{DY} - \varepsilon_{J/\Psi}$. As absortion in the
$J/\Psi$ case is more important than in Drell-Yan case, $\gamma>0$. This means
that the ratio (\ref{12}) decreases with $E_T$ (the first derivative is
negative) and the curvature (the second derivative) is positive. In all
calculations of $J/\Psi$ absortion, including destruction by comovers
\cite{10,11}, the tendency for a large $E_T$ saturation occurs, which implies
positive curvature.

There is, however, another posibility for changing the $\nu$ linearity of
events C in (\ref{5}): if a transition to a quark-gluon plasma occurs. In this
case the $J/\Psi$ and $\Psi'$ formation will be prevented \cite{12}. Now,
besides the change in the effective number $\nu$, it is $\alpha_C$ itself that
changes, becoming a function of $\nu$ and vanishing for large values of $\nu$.

If the transition is an abrupt transition at $\nu=\nu^*$ ($\nu^*$ being energy,
nuclei and acceptance dependent) we have,

\begin{eqnarray}
\alpha_{J/\Psi}(\nu)=\alpha_{J/\Psi}\ \ \, & \nu\leq\nu^*\ \ \, \notag \\
\label{13}\\
\alpha_{J/\Psi}(\nu)=0\ \ \, &  \nu>\nu^*\ \ \, \notag
\end{eqnarray}

\noindent
or, making a more reasonable gaussian approximation to (\ref{13}),

\begin{equation}
\alpha_{J/\Psi}(\nu)=\alpha_{J/\Psi} exp(-\nu^2/\nu^{*^2})\ \ \ .
\label{14}
\end{equation}

\noindent
The $J/\Psi$ to DY ratio becomes now:
\begin{equation}
N_{J/\Psi}(E_T)/N_{DY}(E_T) \sim exp(-E_T^2/E_T^{*^2})\ \ \ ,
\label{15}
\end{equation}

\noindent
to be compared to (\ref{12}). There is an essential difference: if the plasma is
produced the curvature in the $E_T$ dependence of the ratio
$N_{J/\Psi}/N_{DY}$ is negative (for $E_T<E_T^*$).

In Fig.3 we show the NA38/50 data \cite{13,14} on S-U and Pb-Pb collisions
without any curve, for \underline{not} guiding the eye. It is clear that the
S-U data can be fitted using (\ref{12}), with
$\gamma\equiv\varepsilon_{DY}-\varepsilon_{J/\Psi}\simeq 0.25$, and that the
last four points of the Pb-Pb $158$ GeV/c data can be fitted using (\ref{15}),
with $E_T^*\simeq 250$ GeV. An interpretation of the Pb-Pb NA50 results,
including the $\Psi'/\Psi$ ratio, in terms of quark-gluon plasma formations
\cite{15,16} is indeed allowed by data.

Coming back to our basic relation (\ref{11}) we can mention another possibility
for distinguishing absortion effects from quark-gluon plasma formation.

If absortion in the form (\ref{11p}) dominates (or if it is absent,
(\ref{11})), we have

\begin{equation}
P_{J/\Psi}^{Abs}(E_T) 
\begin{Sb} 
\gtrsim \\
E_T\to\infty
\end{Sb}
P(E_T)\ \ \ ,
\label{16}
\end{equation}
 
\noindent
while if plasma is formed, (\ref{13}),
\begin{equation}
P_{J/\Psi}^{Plasma}(E_T)
\begin{Sb}
\ll \\
E_T\to\infty
\end{Sb}
P(E_T)\ \ \ ,
\label{17}
\end{equation}

\noindent
where $P(E_T)$ is the normalized minimum bias distribution. The S-U data are
consistent with (\ref{16}), see \cite{9}.

In conclusion, our main relations (\ref{10}) and (\ref{11}) seem to have
reasonable theoretical justification and to work fairly well. They can be used
to test the formation of the plasma of quarks and gluons, inequalities
(\ref{16}) and (\ref{17}). On the other hand, in the direct comparison of
$J/\Psi$ to DY production the sign of the curvature of the $E_T$ dependence of
the ratio $J/\Psi/DY$ may be critical, Eqs. (\ref{12}) and (\ref{15}).

{\bf Acknowledgements}

We thank CICYT of Spain for financial support and one of us (C.A.S.) to
Xunta de Galicia for a fellowship.

\pagebreak

\newpage

\begin{center}
{\bf Figure captions}\\
\end{center}
\vskip 1cm
\noindent
{\bf Fig.1}. NA38 experimental associated multiplicity to Drell-Yan pairs in S-U
collisions (cross points) compared to ${n\over <n>}P(n)$ where $P(n)$
is the experimental multiplicity distribution of S-U
collisions (squared points).

\vskip 1cm
\noindent
{\bf Fig.2}. Same as fig 1 but now the cross points are the experimental 
associate
multiplicity to $J/\Psi$ production and comparison is made with 
${n^\varepsilon\over
<n^\varepsilon>}P(n)$, and $\varepsilon=0.7$.

\vskip 1cm
\noindent
{\bf Fig.3}. Experimental $J/\Psi$ over DY pairs as a function of $E_T$ in S-U
collisions.

\vskip 1cm
\noindent
{\bf Fig.4}. Experimental $J/\Psi$ over DY pairs as a function of $E_T$ in Pb-Pb
collisions.

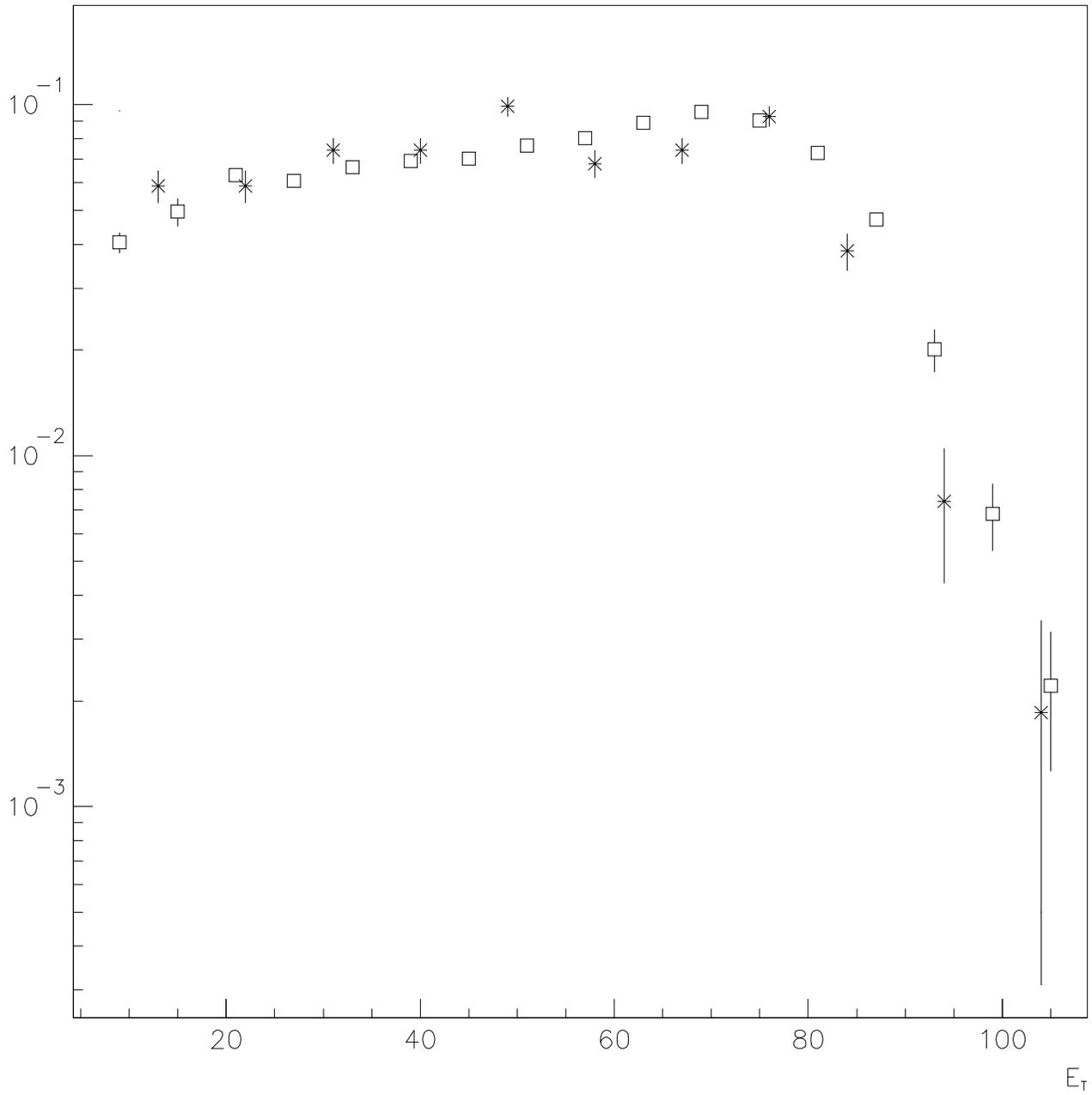
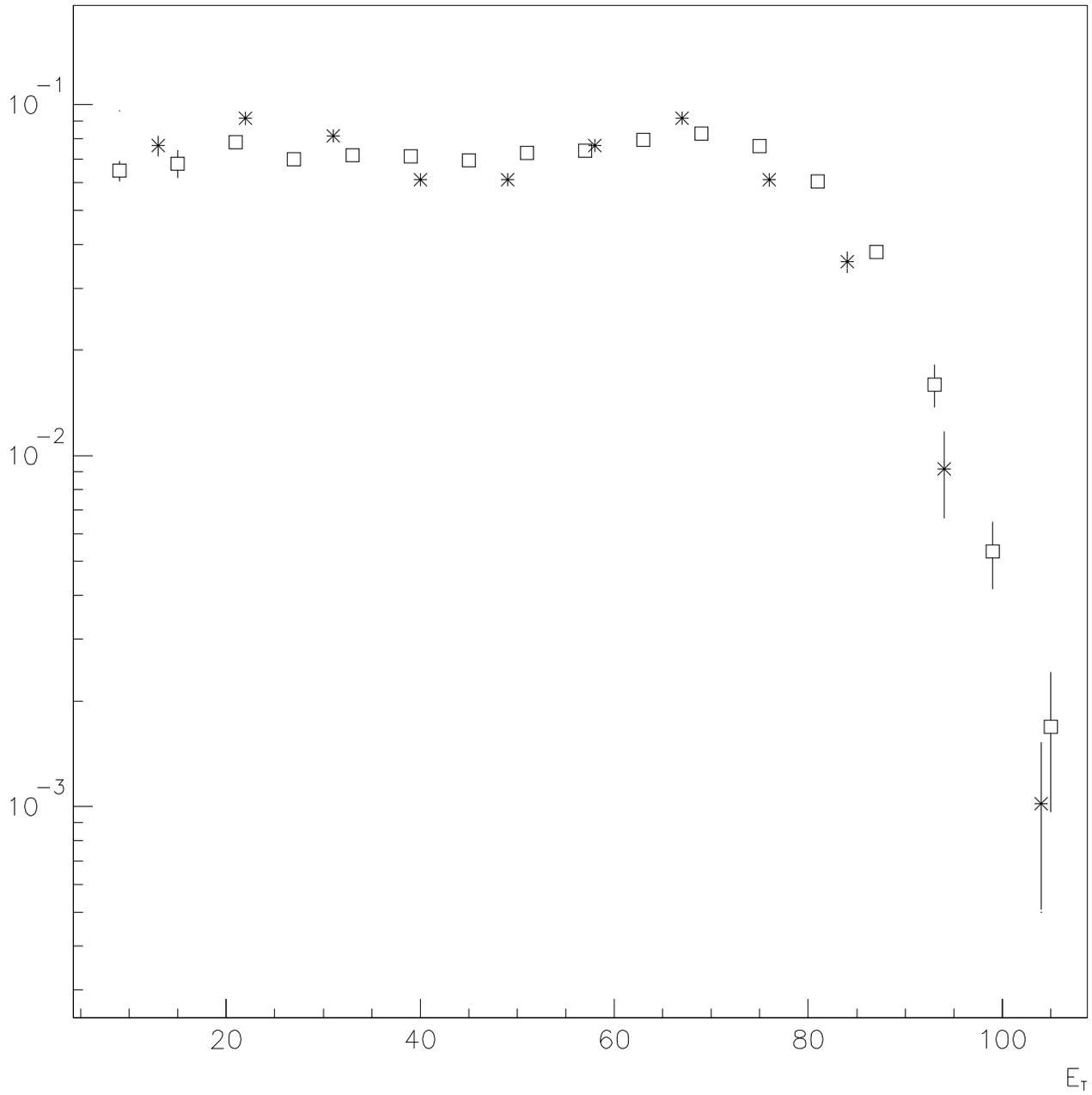
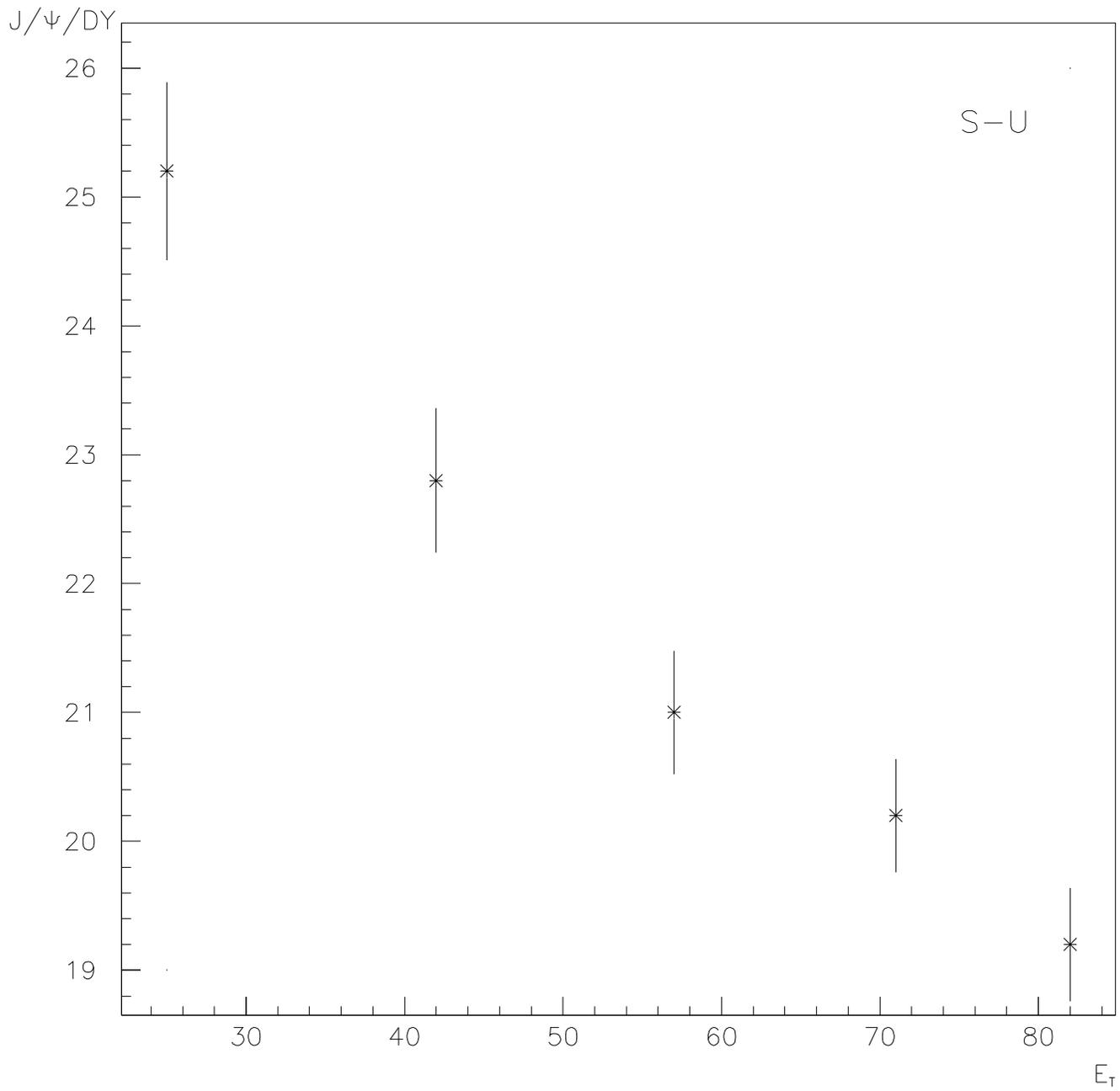
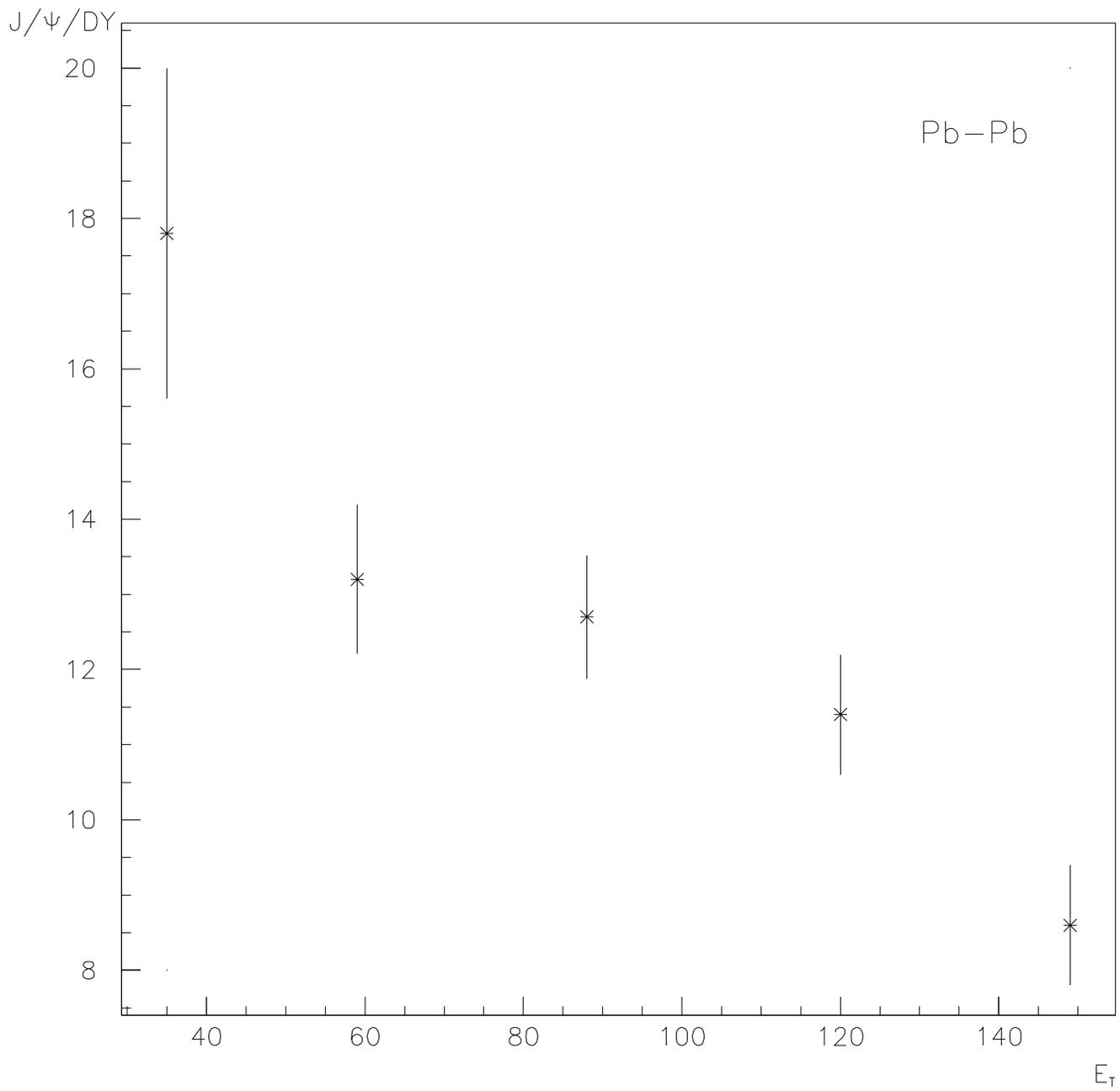
\end{document}